\documentclass{appolb}
\usepackage{graphicx}
\usepackage{physics}
\usepackage{nicefrac}

\newcommand{\as}{{\alpha_S}}

\begin{document}
\title{High-precision prediction for\\ multi-scale processes at the LHC%
\thanks{Presented at Epiphany Conference 2024}
}
\author{Rene Poncelet
\address{Instytut Fizyki Jadrowej Polskiej Akademii Nauk\\ ul. Radzikowskiego 152 31-342 Kraków}
}

\maketitle

\begin{abstract}
Comparisons of higher-order predictions within the Standard Model of Particle Physics (SM) to data are central to high-energy collider experiments like the Large Hadron Collider (LHC). Processes with multiple kinematic scales, such as multi-jet and prompt photon production, provide a unique possibility for probing Quantum Chromodynamics (QCD). These processes directly test perturbative QCD and can be used to extract fundamental parameters like the strong coupling constant and to search for BSM physics. Recent developments enabled lifting three-jet, photon plus two-jet, photon-pair plus jet, and three-photon cross-sections to QCD's next-to-next-to-leading order (NNLO). This contribution presents phenomenological results at NNLO QCD for three-jet and photon plus two-jet production. 

\end{abstract}
  
\section{Introduction}

Tests of perturbative QCD are a staple of the LHC physics program. Processes at high energy allow for comparisons between data and numerical predictions derived from perturbation theory. Calculations performed at leading or next-to-leading perturbative orders (LO and NLO) can readily be obtained from public and automated software, but suffer from large corrections from missing higher orders, mostly estimated through the dependence on non-physical scales. Therefore, computations performed at NNLO or even next-to-next-to-next-to-leading are needed to stabilize the perturbative series such that comparison to data can be performed reliably. For Standard Model processes with low multiplicity, e.g. one or two particles in the final state at leading-order, next-to-next-to-leading order have become the state-of-the-art in the past two decades. This has been driven by the development of subtraction and slicing methods to control the infrared singularities of real emission contributions and improved techniques for loop amplitude calculations. A review of these achievements can be found in Ref.~\cite{Gehrmann:2021qex}.

The methods' extensions to higher multiplicity processes, i.e. processes with more kinematical scales, faced two substantial bottlenecks: Firstly, the computational efficiency and generality of the subtraction and slicing methods had to be improved to deal with the increasing complexity of the phase space and infrared singularity structure. Secondly, the required two-loop amplitudes, whose structures are increasingly more complex, had to be calculated. Recent advances on both fronts have made the computation of NNLO QCD corrections possible for processes with three massless final state particles at the tree level: three-photon production \cite{Chawdhry:2019bji, Abreu:2020cwb, Chawdhry:2020for, Kallweit:2020gcp, Abreu:2023bdp}, diphoton + jet \cite{Agarwal:2021grm, Chawdhry:2021mkw, Agarwal:2021vdh, Chawdhry:2021hkp, Badger:2021imn}, photon + di-jet \cite{Badger:2023mgf} and three-jet production \cite{Abreu:2018jgq, Abreu:2021oya, Czakon:2021mjy, Alvarez:2023fhi} (also the purely gluonic component \cite{Chen:2022ktf}). Also, the first computations of processes where one of the three particles has a non-vanishing invariant mass have been performed \cite{Badger:2021nhg, Badger:2021ega, Abreu:2021asb, Hartanto:2022qhh, Buonocore:2022pqq, Badger:2022ncb}.

This contribution highlights some of the results for three-jet and photon + di-jet production, focusing on the phenomenological results and first comparisons to data.

\section{Computational methods and setup}

The presented calculations have been performed in the formalism of collinear factorisation, where the hadronic cross-section of production of a final state $X$ in the scattering of two hadrons $h_1$ and $h_2$ is written as a convolution of parton distribution functions (PDFs) ($\phi_{i,h}(x,\mu_F)$) with the partonic cross-section $\hat{\sigma}$ for producing the same final state:
\begin{align}
    \sigma_{h_1h_2\to X} = &\sum_{ij} \int_0^1 \text{d}x_1\text{d}x_2 \phi_{i,h_1}(x_1,\mu_F^2)\phi_{j,h_2}(x_2,\mu_F^2)
    \hat{\sigma}_{ij \to X}(\mu^2_R,\mu_F^2)\;.
\end{align}
This holds up to power corrections in  $\nicefrac{\lambda_{\text{QCD}}}{Q}$. Here $\mu_F$ denotes the factorisation and $\mu_R$ the renormalisation scale. The partonic cross-section $\hat{\sigma}$ can be expanded in the strong coupling constant
\begin{align}
    \hat{\sigma}_{ab \to X} = \hat{ \sigma}^{(0)}_{ab \to X}+\as^1 \hat{\sigma}^{(1)}_{ab \to X}+\as^2 \hat{\sigma}^{(2)}_{ab \to X} + \order{\as^3}\;.
\end{align}
The arising infra-red singularities beyond the tree-level approximation in this series cancel in sufficiently inclusive quantities; their treatment in calculations showed in this contribution is done within the Sector-improved residue subtraction scheme \cite{Czakon:2010td, Czakon:2014oma, Czakon:2019tmo} as implemented in the c++ code \textsc{Stripper}. All necessary tree-level amplitudes are evaluated using the \textsc{AvH}-library \cite{Bury:2015dla}. The one-loop amplitudes that start to contribute at the first relative order in $\as$ are evaluated using the \textsc{OpenLoops2} software \cite{Buccioni:2019sur}.

No automated numerical method is currently available for the two-loop amplitudes, starting to contribute at order $\order{\as^2}$. They are, therefore, derived case-by-case in analytical computations. All necessary massless five-point helicity amplitudes have been obtained in a series of publications by different groups, first in the planar or leading-color approximation  \cite{Abreu:2018jgq, Abreu:2020cwb, Chawdhry:2020for, Agarwal:2021grm, Chawdhry:2021mkw} and more recently with the complete colour dependence \cite{Agarwal:2021vdh, Badger:2021imn, Agarwal:2023suw, Abreu:2023bdp, Badger:2023mgf}. The amplitudes are represented in terms of so-called 'Pentagon'-functions \cite{Chicherin:2018old, Chicherin:2020oor, Canko:2020ylt, Abreu:2021smk, Chicherin:2021dyp}, rational coefficient functions of the external kinematic invariants and phase factors. For the phenomenological applications presented here, the amplitudes have been implemented in an independent C++ code. The only exception is the set of amplitudes for three-jet production, where the public code presented in Ref.~\cite{Abreu:2021oya} has been employed. Both implementations allow a stable and fast evaluation of the amplitudes during the integration of the cross-section.

\section{Three-jet production}
Tests of perturbative QCD and measurements of the strong coupling constant using multi-jet rates and event shape observables have a long history. At lepton-colliders, these measurements present the first direct confirmations of the validity of QCD and allow a precise extraction of the fundamental parameters of the theory, i.e. the coupling constant $\as$ and the number of colours. At hadron-hadron colliders, the main challenge in achieving competitive measurements despite the enormous amount of available data is the large theoretical uncertainty in perturbative calculations of the relevant observables. In particular, the dependence on unphysical scales is by far the largest uncertainty in the extraction of $\as$ from jet rates or event-shapes. This section reviews some of the results for the novel NNLO QCD corrections obtained in Ref.~\cite{Czakon:2021mjy, Alvarez:2023fhi} overcoming these bottlenecks. Also, a complete overview of the technical details can be found here.

The basic quantity is the (differential) ratio of the inclusive three-jet production rate to the inclusive dijet-jet production rate:
\begin{align}
    \frac{\dd R_{3/2}(\mu_R,\mu_F)}{\dd X}  = \frac{\dd \sigma_3(\mu_R,\mu_F)/\dd X}{\dd \sigma_2(\mu_R,\mu_F) /\dd X}\,.
\label{eq:ratio_def}
\end{align}
Here, $\dd \sigma_n$ is defined as the (differential) $n$-jet cross-section for having at least $n$ reconstructed anti-kT jets fulfilling analysis-dependent phase space requirements. The renormalisation and factorisation scales in these computations have been set to the scalar sum partonic transverse momenta
\begin{align}
    \mu_R = \mu_F = \hat{H}_T = \sum_{i\in \text{partons}} p_{T,i}\,,
    \label{eq:def_scale}
\end{align}
possibly divided by some constant rational factor. The PDFs are evaluated using the LHAPDF package \cite{Buckley:2014ana}; if not specified otherwise, the NNPDF3.1 PDF \cite{NNPDF:2017mvq} parameterization is used. Estimates of uncertainties from missing higher orders (MHO) are obtained from conventional 7-point scale variations by a factor of $2$, i.e. scale choices within the constraints $\nicefrac{1}{2} \leq \nicefrac{\mu_F}{\mu_R} \leq 2$.

\begin{figure}[htb]
\centerline{
\includegraphics[width=6.25cm,trim=0 -2.5cm 0 0cm]{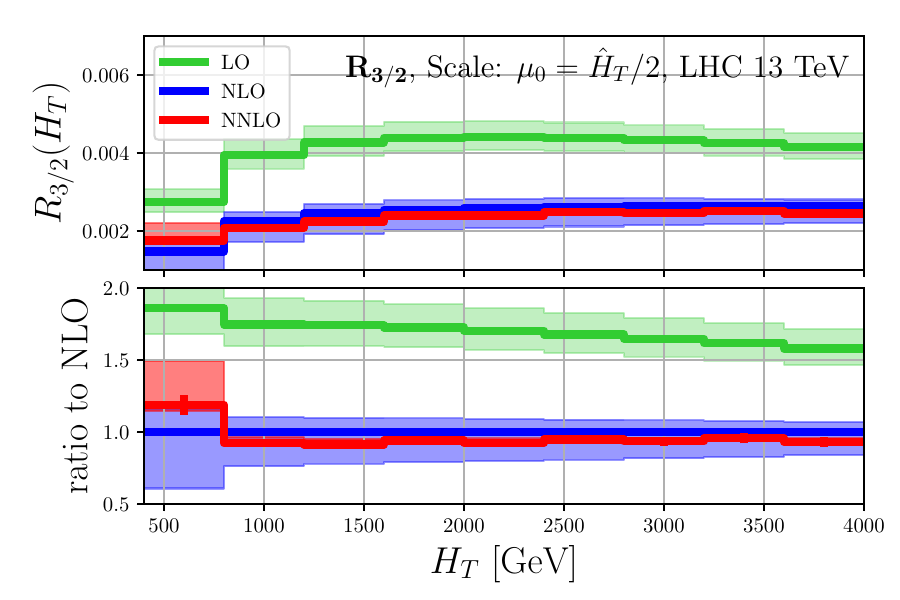}
\includegraphics[width=6.25cm]{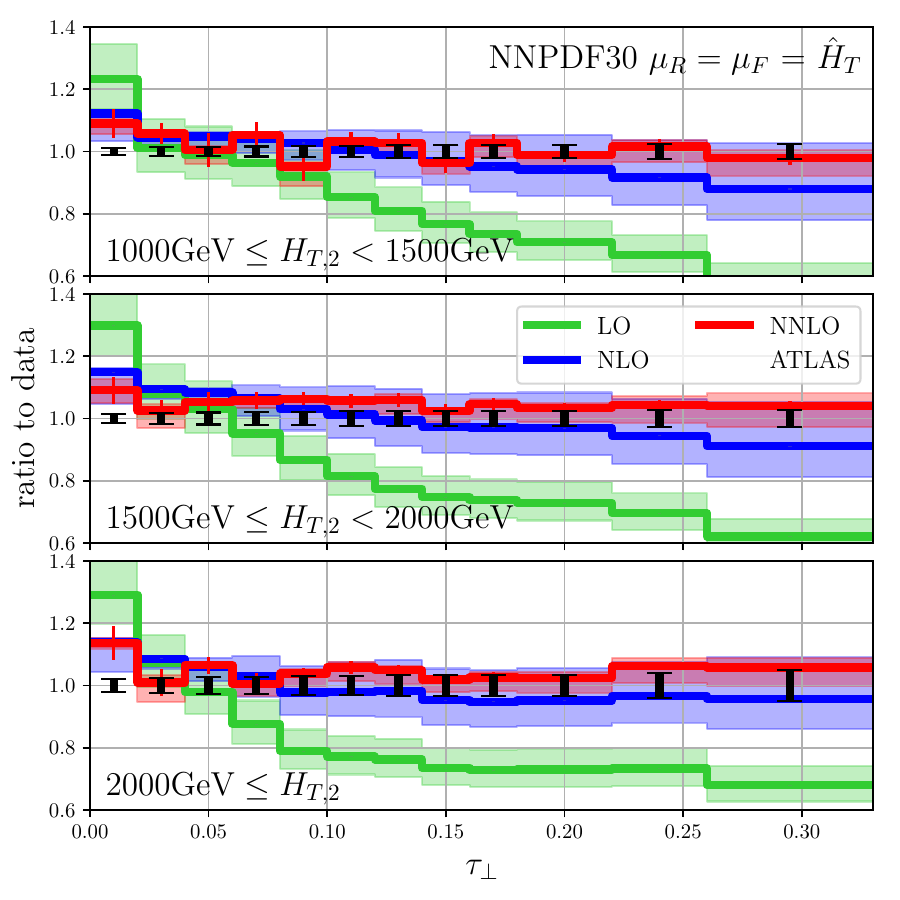}}
\caption{Perturbative predictions through LO (green), NLO (blue) and NNLO (red) QCD at 13 TeV. Bands indicate estimates of corrections from missing higher orders. Left: plot of $\dd R_{3,2} /\dd H_T$, the upper panel shows absolute values, and the lower panel ratios with respect to NLO QCD. Further details in Ref.~\cite{Czakon:2021mjy}. Right: plot of the transverse thrust observable $\tau_{\perp}$ in different regions of $H_{T,2}$ compared to ATLAS data (black)\cite{ATLAS:2020vup}. For details, refer to Ref.~\cite{Alvarez:2023fhi}.}
\label{Fig:3j-r32}
\end{figure}

As an example, the left-hand side in Fig.~\ref{Fig:3j-r32} shows the differential $R_{3/2}$-ratio with respect to $H_T = \sum_{i \in \text{jets}} p_T(j_i)$ for 13 TeV proton-proton collisions. The $\order{1}$ differences between LO and NLO QCD indicate that higher-order corrections are important to describe this observable. At NLO QCD, the MHO uncertainty estimates are of $\order{20\%}$ which cover the actual NNLO QCD corrections, which are, for high values of $H_T$ ($>800 \text{GeV}$), about $3-5\%$. Estimates of corrections from beyond NNLO QCD are tiny and not visibly resolved. The feature in the first bin can be traced back to sensitivity to the phase space boundaries and corresponding enhancements and instabilities.

Event shape observables have been designed to study the geometry of events as a whole, not only a mere sum of its constituents. The (transverse) 'thrust'-observable $T_{\perp}$ (or rather $\tau_{\perp} = 1- T_{\perp}$) \cite{Brandt:1964sa, Farhi:1977sg}, for example, separates isotropic from anisotropic back-to-back configurations and is defined by
\begin{align}
    T_\perp = \max_{\hat{n}_{\perp}} \left\{\frac{\sum_i | \vec{p}_{T,i} \cdot \hat{n}_{\perp}|}{\sum_i | \vec{p}_{T,i}|} \right\}\;,
\label{eq:T-def}    
\end{align}
where the $\hat{n}_{\perp}$ that maximises the expression is called the transverse thrust axis. This quantity (together with several other event shapes) has been measured by the ATLAS collaboration in regions of $H_{T,2} = p_{T,1} + p_{T,2}$ and compared to predictions from Monte Carlo simulations in Ref.~\cite{ATLAS:2020vup}. While the overall description through the simulation is reasonable, several differences in shapes and normalization motivated the inclusion of NNLO QCD corrections in Ref.~\cite{Alvarez:2023fhi}. For example, the right-hand side of Fig.~\ref{Fig:3j-r32} shows perturbative QCD predictions for this observable comparison with ATLAS data. The description of the data improves with increasing perturbative order. Also clearly visible is the significant reduction of the MHO uncertainty estimates at NNLO QCD. The data is fully compatible with these uncertainties (taking into account the remaining statistical uncertainties, shown as vertical bars).

\begin{figure}[htb]
\centerline{%
\includegraphics[width=6.25cm]{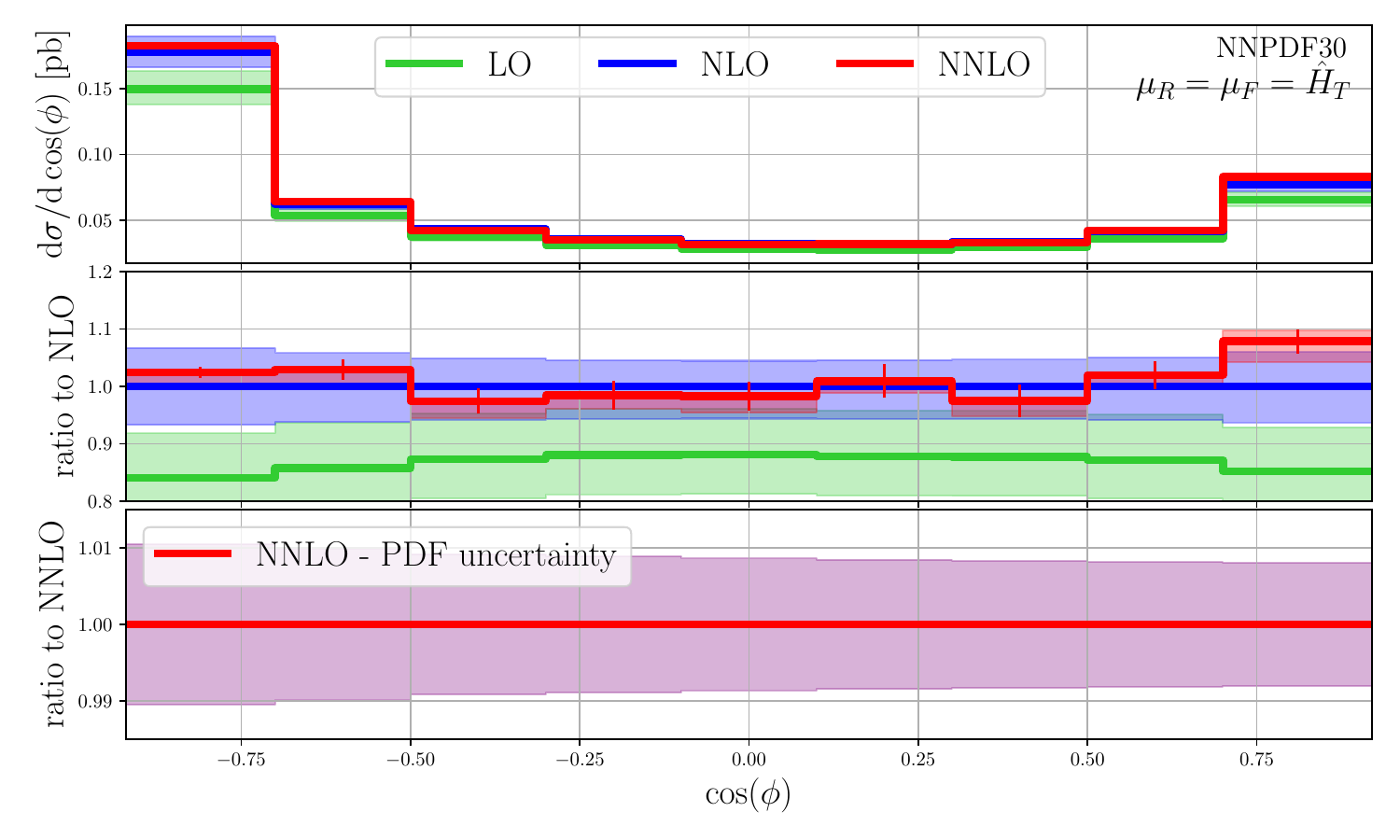}
\includegraphics[width=6.25cm]{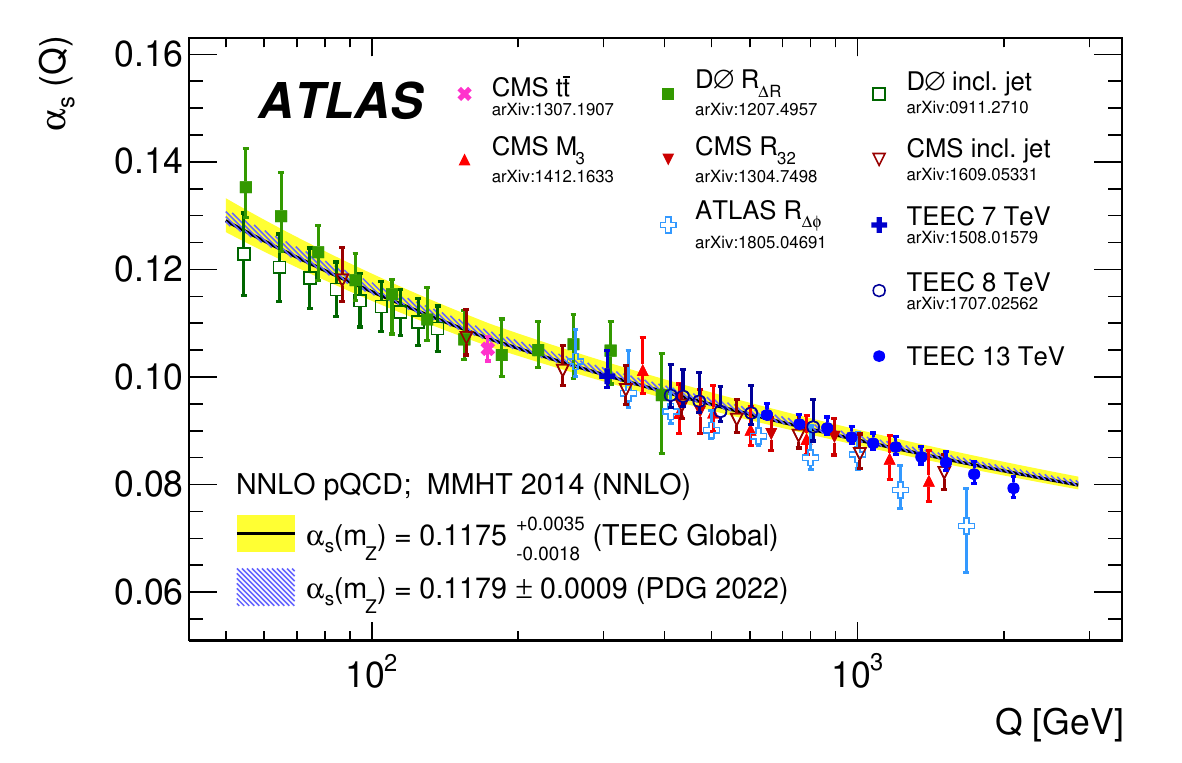}}
\caption{Left: Plots of perturbative predictions (LO - green, NLO - blue, NNLO - red) for the TEEC event shape observable: the upper pane shows the absolute distribution, the central pane the results as a ratio to NLO QCD and the lower pane the PDF uncertainties for reference. For details, see Ref.~\cite{Alvarez:2023fhi}. Right: The extraction of the strong coupling constant as a function of the energy scale, taken from Ref.~\cite{ATLAS:2023tgo}.}
\label{Fig:3j-teec}
\end{figure}

The last highlighted multi-jet observable is the transverse energy-energy correlator (TEEC) \cite{CERN-TH-3800, ATLAS:2023tgo}. Perturbative QCD results obtained in \cite{Alvarez:2023fhi} can be found on the left-hand side of Fig.~\ref{Fig:3j-teec}. Similar to $\tau_T$, a clear reduction of perturbative corrections and MHO uncertainties can be observed when going to NNLO QCD. These results have been used in a refined form in an experimental publication \cite{ATLAS:2023tgo} performing a measurement of the TEEC from multi-jet events to not only make a comparison between theory and data but also to extract $\as$ as a function of the event's energy scale, see the right-hand side in Fig.~\ref{Fig:3j-teec}. Similar extractions have been performed previously using only NLO QCD accuracy but had up to 3 times larger theory uncertainties, which is the dominating uncertainty.

\section{Prompt photon production in association with jets}

The second example of a massless two-to-three process in this contribution is the production of an isolated photon associated with a pair of jets \cite{Badger:2023mgf}. Photon production in hadron-hadron collisions is an important probe of QCD and possesses a rich phenomenology. From the perspective of perturbative QCD, highly energetic prompt photons are of particular interest. There are two main mechanisms to produce prompt photons: fragmentation and direct photon production. To identify prompt photons, isolation criteria, e.g. cuts on the hadronic activity in a cone around candidate photons, are used to suppress background contributions such as those from hadron decays. Unfortunately, these experimental criteria are not infrared-safe in perturbative calculations and require either the inclusion of fragmentation in the calculation or a prescription to remove or suppress these contributions. A simple but phenomenologically effective method is the smooth or hybrid cone isolation \cite{Frixione:1998jh, Siegert:2016bre}. In Ref.~\cite{Badger:2023mgf}, a hybrid cone isolation prescription has been employed. Further details about the calculations, such as phase space definition and observables, can also be found therein.

\begin{figure}[htb]
\centerline{%
\includegraphics[width=6.25cm]{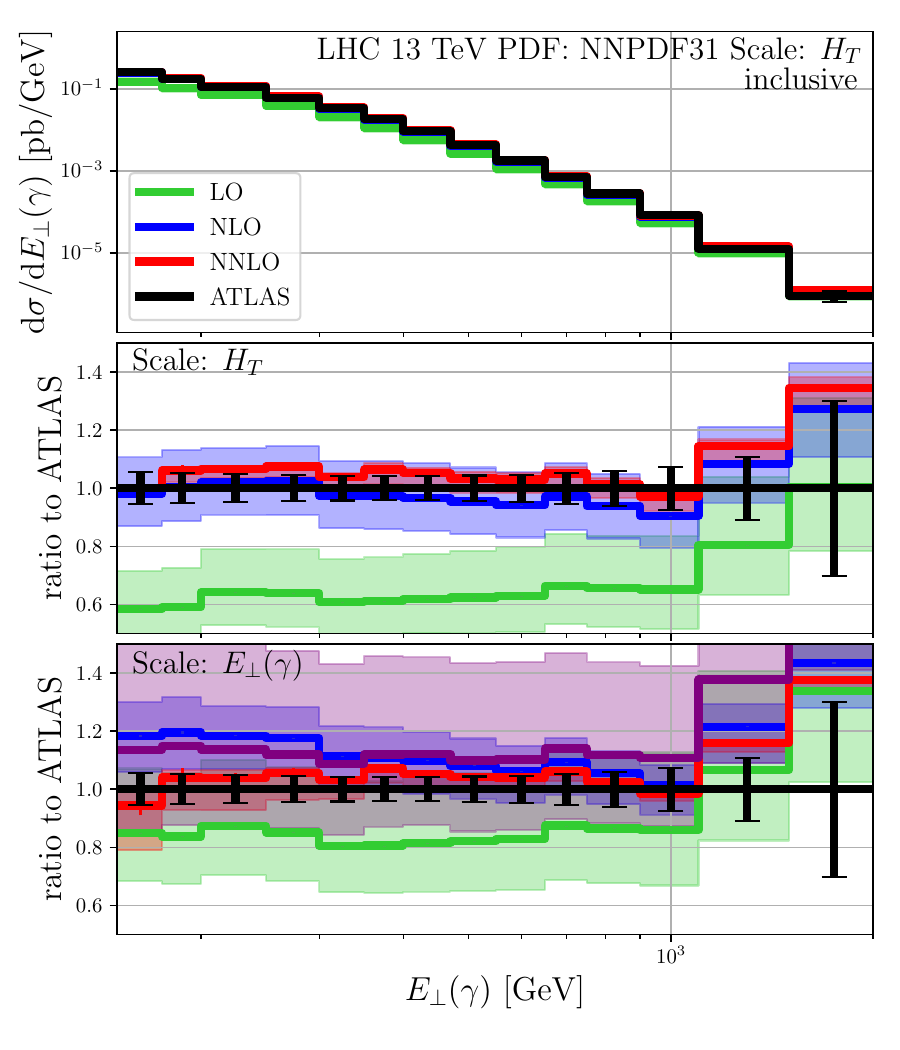}
\includegraphics[width=6.25cm]{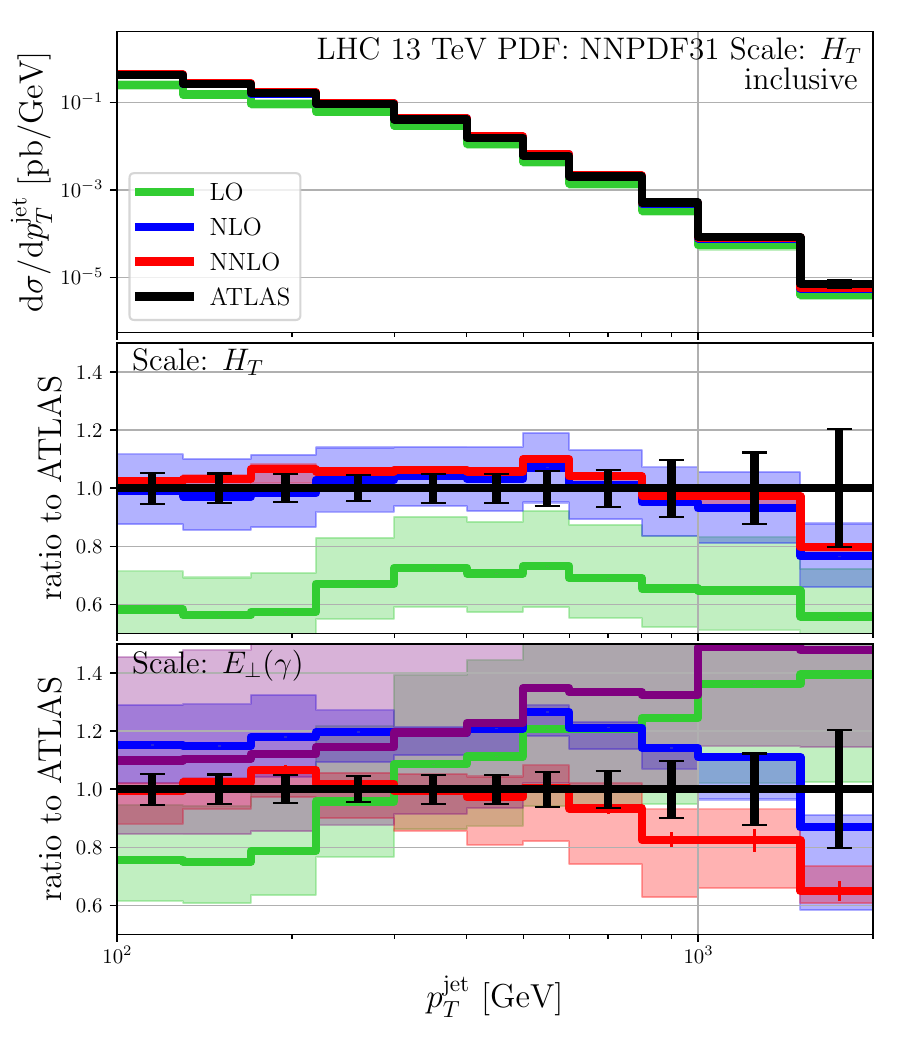}}
\caption{Plot of differential distributions in photon plus di-jet production computed in perturbation theory (LO - green, NLO - blue, NNLO - red), compared to ATLAS data (black) and Monte Carlo predictions (purple). For details see Ref.~\cite{Badger:2023mgf}.}
\label{Fig:a2j-pt}
\end{figure}

Fig.~\ref{Fig:a2j-pt} shows two differential distributions, the transverse energy spectrum of the photon $E_T(\gamma)$ and the transverse momentum distribution of the two leading jets $p^{\text{jet}}_T$ (both jets are accounted for in the same histogram). The perturbative computations are compared to data measured by the ATLAS collaboration \cite{ATLAS:2019iaa}. The perturbative corrections indicate a well-behaving perturbative series, i.e. the corrections are getting smaller, and the scale uncertainties are reduced. The NNLO QCD prediction describes the data well. Besides the perturbative corrections, two different scale choices, $\mu = H_T = E_T(\gamma) + p_T(j_1) + p_T(j_2)$ and $\mu = E_T(\gamma)$ are compared. In both distributions, one observes that the $H_T$ scale has a better perturbative convergence behaviour and, therefore, seems to represent the physical scales much better than $E_T(\gamma)$.

\section{Summary and outlook}
In recent years the computation of second-order QCD corrections to the cross-section of all massless two-to-three processes at the LHC have been performed: $pp \to \gamma\gamma\gamma$ \cite{Chawdhry:2019bji}, $pp \to \gamma \gamma j $ \cite{Chawdhry:2021hkp}, $pp \to \gamma j j $ \cite{Badger:2023mgf} and $pp \to j j j$ \cite{Czakon:2021mjy, Alvarez:2023fhi}. These computations are complete in $n_f = 5$ masslesss QCD as far as the double real and real-virtual corrections are concerned, and have employed a planar/leading-colour approximation for the double virtual contributions. This is except for \cite{Badger:2023mgf}, which, besides being the first computation with complete double virtual amplitudes, demonstrated that this approximation was well justified. The results have been compared to data, and agreement within the uncertainty has been found. The results for three-jet production have been used by the ATLAS collaboration to extract the strong coupling from TEECs for the first time using NNLO QCD accurate theory.

The limitation of the planar approximation for double virtual corrections has been lifted, and all relevant amplitudes are available in complete form. Their inclusion in phenomenological studies is a logical next step. Processes with a single massive (colour-less) particle, such as $V+2j$ processes, are the natural next challenge, for which first steps have been taken \cite{Hartanto:2022qhh, Buonocore:2022pqq}. Further progress is mostly tied to the computation of the respective two-loop amplitudes.

\bibliographystyle{unsrt}
\bibliography{lit}

\begin{thebibliography}{10}

\bibitem{Gehrmann:2021qex}
Thomas Gehrmann and Bogdan Malaescu.
\newblock {Precision QCD Physics at the LHC}.
\newblock {\em Ann. Rev. Nucl. Part. Sci.}, 72:233--258, 2022.

\bibitem{Chawdhry:2019bji}
Herschel~A. Chawdhry, Micha\l{} Czakon, Alexander Mitov, and Rene Poncelet.
\newblock {NNLO QCD corrections to three-photon production at the LHC}.
\newblock {\em JHEP}, 02:057, 2020.

\bibitem{Abreu:2020cwb}
S.~Abreu, B.~Page, E.~Pascual, and V.~Sotnikov.
\newblock {Leading-Color Two-Loop QCD Corrections for Three-Photon Production
  at Hadron Colliders}.
\newblock {\em JHEP}, 01:078, 2021.

\bibitem{Chawdhry:2020for}
Herschel~A. Chawdhry, Micha\l{} Czakon, Alexander Mitov, and Rene Poncelet.
\newblock {Two-loop leading-color helicity amplitudes for three-photon
  production at the LHC}.
\newblock {\em JHEP}, 06:150, 2021.

\bibitem{Kallweit:2020gcp}
Stefan Kallweit, Vasily Sotnikov, and Marius Wiesemann.
\newblock {Triphoton production at hadron colliders in NNLO QCD}.
\newblock {\em Phys. Lett. B}, 812:136013, 2021.

\bibitem{Abreu:2023bdp}
Samuel Abreu, Giuseppe De~Laurentis, Harald Ita, Maximillian Klinkert, Ben
  Page, and Vasily Sotnikov.
\newblock {Two-loop QCD corrections for three-photon production at hadron
  colliders}.
\newblock {\em SciPost Phys.}, 15(4):157, 2023.

\bibitem{Agarwal:2021grm}
Bakul Agarwal, Federico Buccioni, Andreas von Manteuffel, and Lorenzo Tancredi.
\newblock {Two-loop leading colour QCD corrections to $q \bar{q} \to \gamma
  \gamma g$ and $q g \to \gamma \gamma q$}.
\newblock {\em JHEP}, 04:201, 2021.

\bibitem{Chawdhry:2021mkw}
Herschel~A. Chawdhry, Micha\l{} Czakon, Alexander Mitov, and Rene Poncelet.
\newblock {Two-loop leading-colour QCD helicity amplitudes for two-photon plus
  jet production at the LHC}.
\newblock {\em JHEP}, 07:164, 2021.

\bibitem{Agarwal:2021vdh}
Bakul Agarwal, Federico Buccioni, Andreas von Manteuffel, and Lorenzo Tancredi.
\newblock {Two-Loop Helicity Amplitudes for Diphoton Plus Jet Production in
  Full Color}.
\newblock {\em Phys. Rev. Lett.}, 127(26):262001, 2021.

\bibitem{Chawdhry:2021hkp}
Herschel~A. Chawdhry, Micha\l{} Czakon, Alexander Mitov, and Rene Poncelet.
\newblock {NNLO QCD corrections to diphoton production with an additional jet
  at the LHC}.
\newblock {\em JHEP}, 09:093, 2021.

\bibitem{Badger:2021imn}
Simon Badger, Christian Br\o{}nnum-Hansen, Dmitry Chicherin, Thomas Gehrmann,
  Heribertus~Bayu Hartanto, Johannes Henn, Matteo Marcoli, Ryan Moodie, Tiziano
  Peraro, and Simone Zoia.
\newblock {Virtual QCD corrections to gluon-initiated diphoton plus jet
  production at hadron colliders}.
\newblock {\em JHEP}, 11:083, 2021.

\bibitem{Badger:2023mgf}
Simon Badger, Micha\l{} Czakon, Heribertus~Bayu Hartanto, Ryan Moodie, Tiziano
  Peraro, Rene Poncelet, and Simone Zoia.
\newblock {Isolated photon production in association with a jet pair through
  next-to-next-to-leading order in QCD}.
\newblock {\em JHEP}, 10:071, 2023.

\bibitem{Abreu:2018jgq}
S.~Abreu, F.~Febres~Cordero, H.~Ita, B.~Page, and V.~Sotnikov.
\newblock {Planar Two-Loop Five-Parton Amplitudes from Numerical Unitarity}.
\newblock {\em JHEP}, 11:116, 2018.

\bibitem{Abreu:2021oya}
S.~Abreu, F.~Febres~Cordero, H.~Ita, B.~Page, and V.~Sotnikov.
\newblock {Leading-color two-loop QCD corrections for three-jet production at
  hadron colliders}.
\newblock {\em JHEP}, 07:095, 2021.

\bibitem{Czakon:2021mjy}
Micha\l{} Czakon, Alexander Mitov, and Rene Poncelet.
\newblock {Next-to-Next-to-Leading Order Study of Three-Jet Production at the
  LHC}.
\newblock {\em Phys. Rev. Lett.}, 127(15):152001, 2021.
\newblock [Erratum: Phys.Rev.Lett. 129, 119901 (2022), Erratum: Phys.Rev.Lett.
  129, 119901 (2022)].

\bibitem{Alvarez:2023fhi}
Manuel Alvarez, Josu Cantero, Micha\l{} Czakon, Javier Llorente, Alexander
  Mitov, and Rene Poncelet.
\newblock {NNLO QCD corrections to event shapes at the LHC}.
\newblock {\em JHEP}, 03:129, 2023.

\bibitem{Chen:2022ktf}
Xuan Chen, Thomas Gehrmann, E.~W.~N. Glover, Alexander Huss, and Matteo
  Marcoli.
\newblock {Automation of antenna subtraction in colour space: gluonic
  processes}.
\newblock {\em JHEP}, 10:099, 2022.

\bibitem{Badger:2021nhg}
Simon Badger, Heribertus~Bayu Hartanto, and Simone Zoia.
\newblock {Two-Loop QCD Corrections to Wbb\textasciimacron{} Production at
  Hadron Colliders}.
\newblock {\em Phys. Rev. Lett.}, 127(1):012001, 2021.

\bibitem{Badger:2021ega}
Simon Badger, Heribertus~Bayu Hartanto, Jakub Kry\'s, and Simone Zoia.
\newblock {Two-loop leading-colour QCD helicity amplitudes for Higgs boson
  production in association with a bottom-quark pair at the LHC}.
\newblock {\em JHEP}, 11:012, 2021.

\bibitem{Abreu:2021asb}
S.~Abreu, F.~Febres~Cordero, H.~Ita, M.~Klinkert, B.~Page, and V.~Sotnikov.
\newblock {Leading-color two-loop amplitudes for four partons and a W boson in
  QCD}.
\newblock {\em JHEP}, 04:042, 2022.

\bibitem{Hartanto:2022qhh}
Heribertus~Bayu Hartanto, Rene Poncelet, Andrei Popescu, and Simone Zoia.
\newblock {Next-to-next-to-leading order QCD corrections to
  Wbb\textasciimacron{} production at the LHC}.
\newblock {\em Phys. Rev. D}, 106(7):074016, 2022.

\bibitem{Buonocore:2022pqq}
Luca Buonocore, Simone Devoto, Stefan Kallweit, Javier Mazzitelli, Luca
  Rottoli, and Chiara Savoini.
\newblock {Associated production of a W boson and massive bottom quarks at
  next-to-next-to-leading order in QCD}.
\newblock {\em Phys. Rev. D}, 107(7):074032, 2023.

\bibitem{Badger:2022ncb}
Simon Badger, Heribertus~Bayu Hartanto, Jakub Kry\'s, and Simone Zoia.
\newblock {Two-loop leading colour helicity amplitudes for
  W$^{\pm}$\ensuremath{\gamma} + j production at the LHC}.
\newblock {\em JHEP}, 05:035, 2022.

\bibitem{Czakon:2010td}
M.~Czakon.
\newblock {A novel subtraction scheme for double-real radiation at NNLO}.
\newblock {\em Phys. Lett. B}, 693:259--268, 2010.

\bibitem{Czakon:2014oma}
M.~Czakon and D.~Heymes.
\newblock {Four-dimensional formulation of the sector-improved residue
  subtraction scheme}.
\newblock {\em Nucl. Phys. B}, 890:152--227, 2014.

\bibitem{Czakon:2019tmo}
Micha\l{} Czakon, Andreas van Hameren, Alexander Mitov, and Rene Poncelet.
\newblock {Single-jet inclusive rates with exact color at $ \mathcal{O} $ ($
  {\alpha}_s^4 $)}.
\newblock {\em JHEP}, 10:262, 2019.

\bibitem{Bury:2015dla}
M.~Bury and A.~van Hameren.
\newblock {Numerical evaluation of multi-gluon amplitudes for High Energy
  Factorization}.
\newblock {\em Comput. Phys. Commun.}, 196:592--598, 2015.

\bibitem{Buccioni:2019sur}
Federico Buccioni, Jean-Nicolas Lang, Jonas~M. Lindert, Philipp Maierh\"ofer,
  Stefano Pozzorini, Hantian Zhang, and Max~F. Zoller.
\newblock {OpenLoops 2}.
\newblock {\em Eur. Phys. J. C}, 79(10):866, 2019.

\bibitem{Agarwal:2023suw}
Bakul Agarwal, Federico Buccioni, Federica Devoto, Giulio Gambuti, Andreas von
  Manteuffel, and Lorenzo Tancredi.
\newblock {Five-Parton Scattering in QCD at Two Loops}.
\newblock 11 2023.

\bibitem{Chicherin:2018old}
D.~Chicherin, T.~Gehrmann, J.~M. Henn, P.~Wasser, Y.~Zhang, and S.~Zoia.
\newblock {All Master Integrals for Three-Jet Production at
  Next-to-Next-to-Leading Order}.
\newblock {\em Phys. Rev. Lett.}, 123(4):041603, 2019.

\bibitem{Chicherin:2020oor}
Dmitry Chicherin and Vasily Sotnikov.
\newblock {Pentagon Functions for Scattering of Five Massless Particles}.
\newblock {\em JHEP}, 20:167, 2020.

\bibitem{Canko:2020ylt}
Dhimiter~D. Canko, Costas~G. Papadopoulos, and Nikolaos Syrrakos.
\newblock {Analytic representation of all planar two-loop five-point Master
  Integrals with one off-shell leg}.
\newblock {\em JHEP}, 01:199, 2021.

\bibitem{Abreu:2021smk}
Samuel Abreu, Harald Ita, Ben Page, and Wladimir Tschernow.
\newblock {Two-loop hexa-box integrals for non-planar five-point one-mass
  processes}.
\newblock {\em JHEP}, 03:182, 2022.

\bibitem{Chicherin:2021dyp}
Dmitry Chicherin, Vasily Sotnikov, and Simone Zoia.
\newblock {Pentagon functions for one-mass planar scattering amplitudes}.
\newblock {\em JHEP}, 01:096, 2022.

\bibitem{Buckley:2014ana}
Andy Buckley, James Ferrando, Stephen Lloyd, Karl Nordstr\"om, Ben Page, Martin
  R\"ufenacht, Marek Sch\"onherr, and Graeme Watt.
\newblock {LHAPDF6: parton density access in the LHC precision era}.
\newblock {\em Eur. Phys. J. C}, 75:132, 2015.

\bibitem{NNPDF:2017mvq}
Richard~D. Ball et~al.
\newblock {Parton distributions from high-precision collider data}.
\newblock {\em Eur. Phys. J. C}, 77(10):663, 2017.

\bibitem{ATLAS:2020vup}
Georges Aad et~al.
\newblock {Measurement of hadronic event shapes in high-p$_{T}$ multijet final
  states at $ \sqrt{s} $ = 13 TeV with the ATLAS detector}.
\newblock {\em JHEP}, 01:188, 2021.
\newblock [Erratum: JHEP 12, 053 (2021)].

\bibitem{Brandt:1964sa}
S.~Brandt, C.~Peyrou, R.~Sosnowski, and A.~Wroblewski.
\newblock {The Principal axis of jets. An Attempt to analyze high-energy
  collisions as two-body processes}.
\newblock {\em Phys. Lett.}, 12:57--61, 1964.

\bibitem{Farhi:1977sg}
Edward Farhi.
\newblock {A QCD Test for Jets}.
\newblock {\em Phys. Rev. Lett.}, 39:1587--1588, 1977.

\bibitem{CERN-TH-3800}
{A. Ali, E. Pietarinen, W. J. Stirling}.
\newblock {Transverse energy-energy correlations: A test of perturbative QCD
  for the proton-antiproton collider}.
\newblock {\em Phys. Lett. B}, 141:447, 1984.

\bibitem{ATLAS:2023tgo}
Georges Aad et~al.
\newblock {Determination of the strong coupling constant from transverse
  energy$-$energy correlations in multijet events at $\sqrt{s} = 13$ TeV with
  the ATLAS detector}.
\newblock {\em JHEP}, 07:085, 2023.

\bibitem{Frixione:1998jh}
Stefano Frixione.
\newblock {Isolated photons in perturbative QCD}.
\newblock {\em Phys. Lett. B}, 429:369--374, 1998.

\bibitem{Siegert:2016bre}
Frank Siegert.
\newblock {A practical guide to event generation for prompt photon production
  with Sherpa}.
\newblock {\em J. Phys. G}, 44(4):044007, 2017.

\bibitem{ATLAS:2019iaa}
Georges Aad et~al.
\newblock {Measurement of isolated-photon plus two-jet production in $pp$
  collisions at $\sqrt s=13$ TeV with the ATLAS detector}.
\newblock {\em JHEP}, 03:179, 2020.

\end{thebibliography}

\end{document}